\documentclass[dvipsnames]{article}

\usepackage{amsmath}
\usepackage{commath}
\usepackage{url}
\usepackage{cite}
\usepackage{xcolor}
\usepackage{multirow}
\usepackage[margin=1.25in]{geometry}
\usepackage{graphicx}
\graphicspath{{./graphics/}}
\usepackage[font=small,margin=1cm]{caption}
\usepackage{pdfpages}
\usepackage{amssymb}
\usepackage{titling}
\usepackage{lineno}
\usepackage{hyperref} 
\usepackage{float}

\title{Threshold $J/\psi$ Photoproduction as a Probe of Nuclear Gluon Structure}

\author{
V.~Kakoyan, H.~Marukyan, A.~Shahinyan, H.~Voskanyan\\
A.~I.~Alikhanyan National Science Laboratory (Yerevan Physics Institute),\\ 0036 Yerevan, Armenia\\
\and
H.~Gao (spokesperson), Y.~Liu, B.~Yu, Y.~Yu,\\ Z.~Zhao, J.~Zhou\\
Duke University, Durham, North Carolina 27708, USA\\
\and
S.~Paul, H.~Szumila-Vance (spokesperson)\\
Florida International University, Miami, Florida 33199, USA\\
\and
S.~Bhattarai, S.~Regmi\\
Idaho State University, Pocatello, Idaho 83209, USA\\
\and
A.~Hobart\\
IJCLab, CNRS/IN2P3, Université Paris-Saclay, 91405 Orsay, France\\
\and
S.~Fang, B.~Liu, X.~Shen\\
Institute of High Energy Physics, Chinese Academy of Sciences, Beijing 100049, China\\
\and
T.~Kutz\\
Johannes Gutenberg-Universität Mainz, Mainz 55122, Germany\\
\and 
T.~Kolar (spokesperson)\\
Jožef Stefan Institute, Ljubljana 1000, Slovenia\\
\and
J.~R.~Pybus (spokesperson)\\
Los Alamos National Laboratory, Los Alamos, New Mexico 87545, USA\\
\and
L.~Ehinger, O.~Hen (spokesperson), C.-W.~Lin, J.~Phelan,\\ H.~Qi, N.~Wright\\
Massachusetts Institute of Technology, Cambridge, Massachusetts 02139, USA\\
\and
B.~R.~Devkota, D.~Dutta (spokesperson), L.~El Fassi, M.~Maynes,\\ M.~Ouillon, B.~Tamang, U.~Weerasinghe\\
Mississippi State University, Mississippi State, Mississippi 39762, USA\\
\and
S.~Somov\\
National Research Nuclear University MEPhI, Moscow 115409, Russia\\
\and
M.~Paolone\\
New Mexico State University, Las Cruces, New Mexico 88003, USA\\
\and
C.~Ayerbe Gayoso, M.~Hattawy, L.~B.~Weinstein\\
Old Dominion University, Norfolk, Virgnia 23529, USA\\
\and
B.-Q.~Ma\\
Peking University, Beijing 100871, China \\
Zhengzhou University, Henan 450001, China\\
\and
E.~O.~Cohen, J.~Lichtenstadt, I.~Korover (spokesperson), I.~Parshkin,\\ E.~Piasetzky, R.~Wagner\\
Tel Aviv University, Tel Aviv 69978, Israel\\
\and
N.~Sparveris\\
Temple University, Philadelphia, Pennsylvania 19122, USA\\
\and
O.~Cortes Becerra, A.~Schmidt (spokesperson), I.~Strakovsky\\
The George Washington University, Washington, D.C. 20052, USA\\
\and 
H.~Lu, D.~Nguyen\\
The University of Tennessee, Knoxville, TN 37996, USA\\
\and
X.~Bai, V.~V.~Berdnikov, K.~Dehmelt, A.~Deur,\\ J.-O.~Hansen, F.~Hauenstein, D.~Higinbotham, I.~Larin,\\ M.~D.~McCaughan, C.~Morean, A.~Panta, D.~Romanov,\\ D.~Smith, A.~Somov (spokesperson), A.~Tadepalli,\\ R.~Tyson, B.~Zihlmann\\
Thomas Jefferson National Accelerator Facility, Newport News, VA 23606, USA\\
\and
V.~Lyubovitskij\\
Tomsk State University, 634050 Tomsk, Russia\\
\and
Z.~Chen, Z.~Huang, Y.~Li, Z.~Ye,\\Y.~Zhang, Z.~Zhang, H.~Zhao\\
Tsinghua University, Beijing 100084, China\\
\and
R.~Jones\\
University of Connecticut, Storrs 06269, CT, USA\\
\and
B.~McKinnon\\
University of Glasgow, Glasgow G12 8QQ, United Kingdom\\
\and
T.~Xiao\\
University of North Texas, Denton, Texas 76203, USA\\
\and
M.~Junaid\\
University of Regina, Regina, SK S4S 0A2, Canada\\
\and
P.~Gautam\\
University of Virginia, Charlottesville, Virginia 22904, USA\\
\and
D.~Androić\\
University of Zagreb, 10000 Zagreb, Croatia\\
\and
B.~Pandey\\
Virginia Military Institute, Lexington, Virginia 24450, USA\\
\and
N.~Kalantarians\\
Virginia Union University, Richmond, Virginia 23220, USA\\
\and
Z.~Zhang, X.~Zhou\\
Wuhan University, Wuhan, Hubei 430072, China\\
}

\date{\vspace{-5ex}}

\begin{document}

\maketitle

\pagebreak

\section*{Executive Summary}

We summarize here the goals and requirements of the experiment.

{\bf Physics goals.} The primary goal of this experiment is to set the first stringent constraints on the possible modification of gluon behavior at large $x$ in bound nucleons, similarly to the quark-sector ``EMC effect''.
This will be done by studying the incoherent photoproduction of $J/\psi$ from $^{4}$He, with particular emphasis on sub-threshold photoproduction and production from hight-momentum SRC nucleons.

{\bf Proposed Measurements.} The key measurements in this experiment will be the semi-inclusive process $^4$He$(\gamma,J/\psi p)X$; measurement of the total cross section as a function of $E_\gamma$ across the tagged energy range, including below threshold; characterization of $|t|$- and $p_{miss}$-dependence of the cross section.

{\bf Run Conditions.} The experiment will run in Hall D with the standard GlueX configuration. The experiment will require a total of 85 PAC days: 80 days on a $^4$He target and 5 days on $^2$H. 
An electron beam energy of 12 GeV is required to generate the needed coherent photon energy peak of 8 GeV.

\pagebreak

\begin{abstract}
The nuclear EMC effect is the observation that quark distributions in bound nucleons experience significant modification at large $x$ relative to free nucleons. 
Despite decades of measurements verifying the presence of this effect in quarks across a wide range of nuclei, behavior of large-$x$ gluons in nuclei remains almost completely unknown.
As the nuclear physics community seeks out new observables to try to elucidate the mechanisms behind the EMC effect, it becomes striking that we remain ignorant regarding the impact of nuclear effects on gluonic behavior.

Recent photonuclear data using the Hall D photon beam have enabled the first measurement of $J/\psi$ photoproduction from nuclei near and below the energy threshold, with the results highlighted in Physical Review Letters as an Editors' Suggestion. 
These data have placed the first, and currently only, constraints on the behavior of large-$x$ gluons within bound nucleons.
However, compared to the quantity of data which currently informs our knowledge of the quark-sector EMC effect, these data are extremely limited, and remain unable to conclusively observe or exclude large modification of gluon distributions.

A high-luminosity photonuclear experiment will enable a precision measurement of incoherent $J/\psi$ photoproduction at and below the threshold region.
This data will provide the first stringent constraints on nuclear modification of gluon structure or other exotic effects which could impact the production of $J/\psi$ from nuclei.

We request 85 PAC days at Hall D using the GlueX detector with a 12 GeV electron beam energy and a coherent photon peak energy of $8$ GeV, split into 80 days using a $^4$He target and 5 calibration days using a $^2$H target.
\end{abstract}

\pagebreak

\tableofcontents

\pagebreak

\section{Introduction}

The EMC effect~\cite{AUBERT1983275,Malace:2014uea} represents an outstanding puzzle in nuclear physics: quarks in bound nucleons behave differently than for free nucleons, with nuclear structure functions at large-$x$ showing large suppression relative to deuterium which cannot be explained by conventional nuclear physics.
Four decades of deep-inelastic scattering experiments have since confirmed this observation across a wide range of nuclei, including a variety of experiments conducted at Jefferson Lab aiming to finally address this puzzle.
Despite hundreds of papers written on this topic, the exact mechanisms by which bound nucleons are modified remain largely elusive, with the observation that the EMC effect correlates strongly with the abundance of short-range correlated nucleon pairs~\cite{Hen:2016kwk,Weinstein:2010rt,Hen:2012fm,Hen:2013oha} providing some of the only hints at the cause of this effect.

Four decades of experiments have given us a wealth of data on the modification of quark behavior in nuclei at large $x$.
But quarks contribute to only half of nuclear matter, with the other half resulting from the dynamics of the gluons.
In stark contrast to the situation with quark distributions, the behavior of gluons within nuclei at large-$x$ remains almost completely unstudied.
Gluon PDFs in the valence region have extremely few data constraints, and it is presently unknown what modification, if any, they experience in parallel to the quarks.

$J/\psi$ production at JLab has enabled measurement of the gluonic structure of the proton at high-$x$, in the threshold region for photoproduction~\cite{Ali_2019,HallCJPsi}.
Measurements of incoherent $J/\psi$ photoproduction from nuclei have been performed by looking at data from Ultra-Peripheral Collisions at RHIC and the LHC~\cite{PHENIX:2009xtn,Abelev_2013,Abbas_2013} as well as high-energy photon scattering at Fermilab~\cite{PhysRevLett.57.3003}, but these data are statistically limited and probe much lower values of $x$ than photoproduction measurements at JLab.
A precision measurement of incoherent $J/\psi$ production from nuclei with photon energies between $6$ and $12$ GeV would be the first such measurement in the threshold region, and would extend to being the first precise measurement of sub-threshold photoproduction of $J/\psi$.
These measurements can provide the first insights into the gluonic structure of nuclei and bound nucleons at high-$x$.
Such a measurement is a necessary complement to the extensive measurements of the EMC effect.
Measurement of $J/\psi$ from nuclei near and below threshold can give the first constraints on the distributions of gluons within nuclei at similar values of $x$.
Furthermore, the quasi-exclusive nature of incoherent $(\gamma,J/\psi\ p)$ photoproduction allows reconstruction of the initial proton involved in the reaction, allowing for direct testing of the differences between $J/\psi$ photoproduction on mean-field and SRC nucleons.

We have completed the first analysis of near- and sub-threshold photoproduction of $J/\psi$ from nuclear targets $^2$H, $^4$He, and $^{12}$C, recently highlighted as an Editors' Suggested article in PRL~\cite{pybus2024measurementnearsubthresholdjpsi}. The results of this analysis establish the feasibility of such measurements in Hall D and motivate the probe of high-$x$ gluons in the nucleus proposed here.

We propose here an 80-PAC day measurement of the nuclear target $^4$He, using the Hall D real photon beam with a coherent peak energy of $8$ GeV and the GlueX detector in its standard configuration, as well as a 5-PAC day calibration period using $^2$H, with the goal of performing a precision measurement of $J/\psi$ nuclear photoproduction over the near- and sub-threshold energy region.
This beam time will allow a measurement of the total sub-threshold $J/\psi$ photoproduction cross section to an estimated  precision of 35\% (or two energy bins with 50\% precision), which is the target goal of this experiment, as well as a total accumulated 600 $(\gamma,J/\psi p)$ events.

We present an overview of the recent experimental and theoretical results on the EMC effect and threshold $J/\psi$ photoproduction in Section~\ref{sec:recent}.
In Section~\ref{sec:goals}, we outline the primary physics goals of this experiments, and in Section~\ref{sec:measurement}, we detail the proposed experiment, including kinematics of the measurement, optimization of the coherent peak of photon energy, and expected rates for the channels of interest.

\section{Physics Background}
\label{sec:recent}

\subsection{EMC Effect}

The EMC effect was first observed by the EMC collaboration, which measured the structure functions of deuterium and iron in muon deep-inelastic scattering and observed that the structure function ratio $F_2^A/F_2^D$ was significantly depleted in the valence region $0.3\lesssim x\lesssim 0.7$~\cite{AUBERT1983275}.
Since this, measurements of nuclear structure functions have validated this effect across a wide range of nuclei, observing deviations from unity that cannot be fully explained by the effects of nuclear binding or Fermi motion.
Thus far, the abundance of data measuring this effect has not resulted in a clear understanding of the mechanisms producing such a modification of bound nucleon partonic structure, and the EMC effect remains the subject of significant scrutiny and scholarly debate.

A key development in recent years has been the observation that the strength of the EMC effect correlates with the abundance of Short-Range Correlated nucleon-nucleon pairs within nuclei from $^3$He to $^{197}$Au~\cite{Hen:2016kwk,Weinstein:2010rt,Hen:2012fm,Hen:2013oha}.
The observation of a correlation between the strength of the EMC effect and the SRC scaling coefficients in nuclei generated new interest in the EMC effect (see e.g. CERN Courier cover paper from May 2013; `Deep in the nucleus: a puzzle revisited'~\cite{cern:courier}) and gave new insight into its possible origin. 
These data suggest that the EMC effect may be primarily driven by the substantial modification of high-momentum, deeply bound nucleons within SRC pairs, rather than a universal modification of all nucleons within a given nucleus.
Several models have been proposed by us and others that attempt to explain the underlying dynamics that drive the EMC effect and its correlation with SRC pair abundances; see a recent review in Ref.~\cite{Hen:2016kwk}.

\begin{figure}[t]
    \centering
    \includegraphics[width = 0.8 \textwidth]{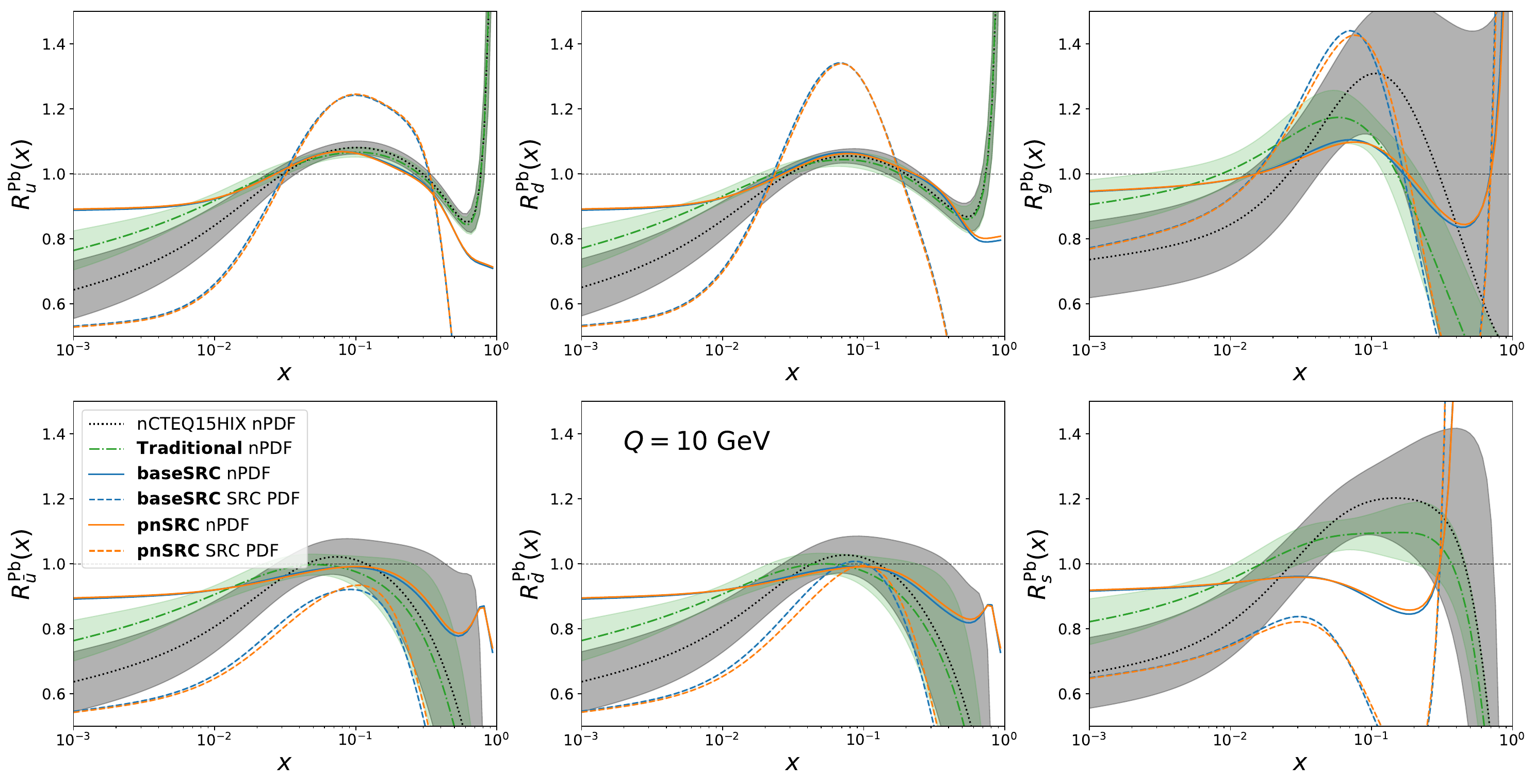}
    \caption{Ratios of nPDFs for $^{208}$Pb to the free-proton PDFs for different parton flavors, from the analysis of Ref.~\cite{2660149}.
    The shaded bands on the full nPDF models show the constraints given the current experimental data. 
    We note in particular the figure on the top-right, which shows the gluon nPDF ratio.
    It can be seen that gluon distributions in nuclei at large $x$ are poorly-constrained at present.}
    \label{fig:nPDF}
\end{figure}

Most recently, a publication by nCTEQ and authors of this proposal expanded this framework to a global analysis of nuclear parton distribution functions (nPDFs)~\cite{2660149}. 
This analysis interpreted data from deep-inelastic scattering, Drell-Yan process, and $W$ and $Z$ boson production simultaneously in a unified framework that incorporated the hypothesis that SRC nucleons alone experience significant modification.
This model was able to match or outperform the traditional parameterizations of nPDFs while using a far more physically-motivated basis that permitted greater interpretation of the results.
Furthermore, the results of the fit allowed for the first extraction of SRC abundances using partonic structure data, giving results which matched previous quasi-elastic scattering data as well as quantum Monte-Carlo calculations.
This result was deemed a 2024 Physics World Breakthrough for advancing our understanding of how the partonic and nucleonic structure of nuclei are related.
Despite the success of this study, a significant limitation in this global analysis was the complete lack of constraints on nuclear gluon PDFs at large $x$, as shown in Fig.~\ref{fig:nPDF}.
This leaves our understanding of the EMC effect fundamentally incomplete.

\subsection{\texorpdfstring{$J/\psi$}{J/psi} Photoproduction}

Photoproduction of the $J/\psi$ meson from the proton was observed at both Cornell~\cite{PhysRevLett.35.1616} ($E_\gamma=11$ GeV) and SLAC~\cite{Camerini:1975cy} ($E_\gamma=19$ GeV) soon after the discovery of the particle, and later at HERMES~\cite{Amarian:1999pi} ($E_\gamma=15$ GeV). 
Since the first observation of the phenomenon $J/\psi$ photoproduction has come to be understood as largely resulting from the exchange of gluons~\cite{Ryskin:1992ui,Brodsky:1994kf,Brodsky_2001}.
The 12-GeV upgrade to Jefferson Lab has enabled the first detailed differential measurements of $J/\psi$ photoproduction near the photoproduction threshold energy of $E_\gamma\approx 8.2$ GeV.

\begin{figure}[t]
    \centering
    \includegraphics[width = 0.48 \textwidth]{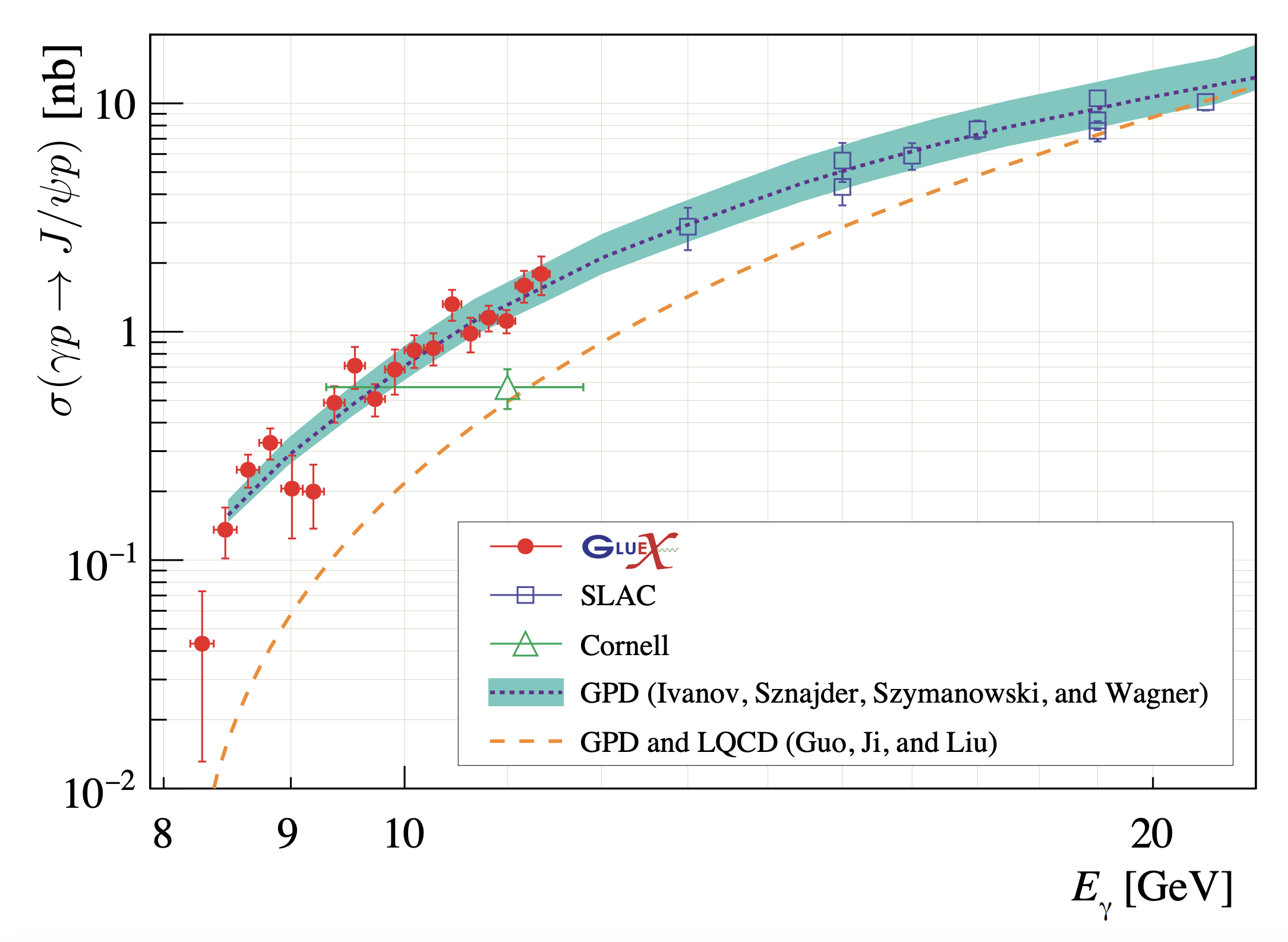}
    \hspace{8mm}
    \includegraphics[width = 0.4 \textwidth]{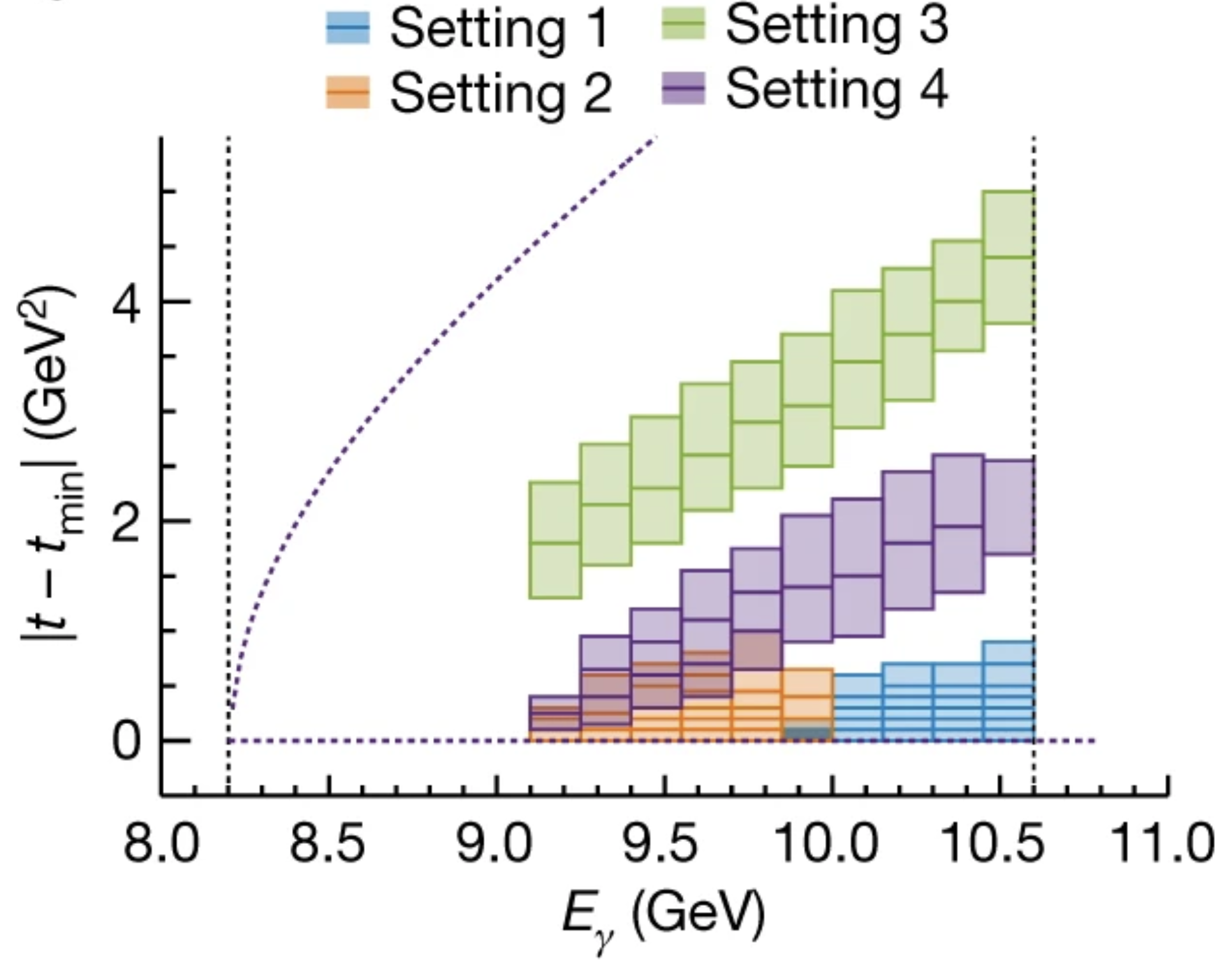}
    \caption{Measurements of $J/\psi$ photoproduction a Jefferson Lab. 
    Left: The total cross section $\sigma(\gamma p\rightarrow J/\psi p)$ as a function of $E_\gamma$ from Ref.~\cite{adhikari2023measurement}.
    Right: Kinematic coverage of the measurement $\frac{d\sigma}{dt}(\gamma p\rightarrow J/\psi p)$ in bins of $E_\gamma$ and $t$ from Ref.~\cite{HallCJPsi}.}
    \label{fig:GlueXJPsi}
\end{figure}

A 2019 study by GlueX~\cite{Ali_2019} used real photon-proton data measured in Hall D to perform the first exclusive measurement of the $\gamma p\rightarrow J/\psi p$ cross section in the threshold region, spanning the photon energy range $8.2<E_\gamma<11.8$ GeV. 
This study, as well as a follow-up study in 2023~\cite{adhikari2023measurement}, measured both the total $J/\psi$ production cross section as a function of photon energy $E_\gamma$ (shown in Fig.~\ref{fig:GlueXJPsi}) and the energy-integrated differential cross section as a function of 4-momentum transfer $t$.

As the first precision $J/\psi$ photoproduction data in the threshold region, this measurement provided substantial new insight into the gluonic structure of the proton not previously possible.
The measurement has enabled insights into the gluonic/mechanical radius of the proton~\cite{PhysRevD.104.054015,PhysRevD.103.L091501}, has been interpreted under the frameworks of gluon Generalized Parton Distributions (GPDs)~\cite{PhysRevD.103.096010} and holographic QCD~\cite{Mamo_2021}, and has aided in understanding the proton mass by allowing extraction of the proton ``trace anomaly'' mass term~\cite{TraceAnomaly}.
The data has additionally enabled extractions of the $J/\psi$-proton scattering length, enabling greater understanding of the formation mechanisms for the $J/\psi$ in photoproduction events~\cite{PhysRevC.101.042201,PhysRevC.101.045201,Du_2020,EPJA-jpsi-scattering,PhysRevD.104.074028,PhysRevC.106.015202}.
A more precise measurement of the process was recently performed using added data from Hall D~\cite{adhikari2023measurement}, enabling improved extrapolation to the forward ($t=0$) differential cross section, an important quantity in many theoretical models of the process~\cite{Kharzeev:1998bz,PhysRevC.101.042201,EPJA-jpsi-scattering,PhysRevD.94.074001}.
These new data also suggest potentially more complicated reaction mechanisms present near threshold, particularly in open-charm box diagrams.

A 2023 study~\cite{HallCJPsi} used real photon-proton data measured in Hall C to perform the first double-differential measurement of $J/\psi$ photoproduction.
While this measurement was not exclusive, detecting only the $J/\psi\rightarrow e^+e^-$ decay, the high luminosity of a spectrometer-based measurement allowed detailed measurements of $\frac{d\sigma}{dt}(\gamma p\rightarrow J/\psi p)$ as a function of both $E_\gamma$ and $t$.
The double-differential nature of this measurement allowed for detailed determination of gluonic gravitational form factors (GFFs) for the proton and higher-precision extraction of the proton trace anomaly mass.
Theoretical analysis of this data was performed using both GPD~\cite{PhysRevD.103.096010} and holographic QCD~\cite{Mamo_2021} and was benchmarked against lattice QCD (LQCD) calculations for the proton~\cite{Pefkou_2022}.
The holographic QCD framework was found to agree particularly well with LQCD predictions, which provides further insight into the reaction mechanisms of $J/\psi$ photoproduction near threshold and enables more precise interpretation of future $J/\psi$ data.

\subsubsection{GlueX-III}

We note that GlueX has recently been approved for an extension ``GlueX-III'', which will be a total of 200 PAC days.
GlueX-III expected to run after the installation of a GEM-TRD detector in the forward region, which will greatly enhance $e/\pi$ separation and improve background rejection in the measurement of $J/\psi\rightarrow e^+e^-$.
This run will be focused on a high-precision measurement of the threshold photoproduction of charmonium, especially the process $\gamma p\rightarrow J/\psi p$.
In addition to providing precise measurements of the free-proton process, which will aid in interpreting the nuclear photoproduction of $J/\psi$, this measurement will help to elucidate outstanding questions regarding the exact mechanisms of $c\bar c$ formation at low energy, which remain the subject of some discussion at present~\cite{adhikari2023measurement}.

\subsection{First Observation of Threshold $J/\psi$ from Nuclei}

The SRC-CT experiment, which measured photonuclear interactions in the GlueX spectrometer for the first time, provided the first opportunity to place constraints on a gluonic EMC effect.
A recent analysis by the authors~\cite{pybus2024measurementnearsubthresholdjpsi} used this data to perform the first measurement of near- and subthreshold photoproduction of $J/\psi$ from nuclear targets $^2$H, $^4$He, and $^{12}$C.
This study provided the first experimental data sensitive to the gluon distributions in bound nucleons at large $x$, a regime unconstrained by previous nPDF data.

\begin{figure}[t]
    \centering
    \includegraphics[width = 0.5 \textwidth]{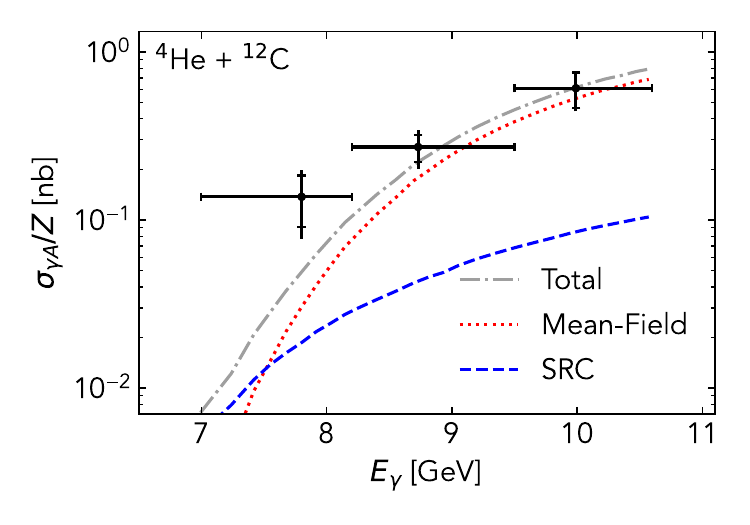}
    \hspace{8mm}
    \includegraphics[width = 0.3 \textwidth]{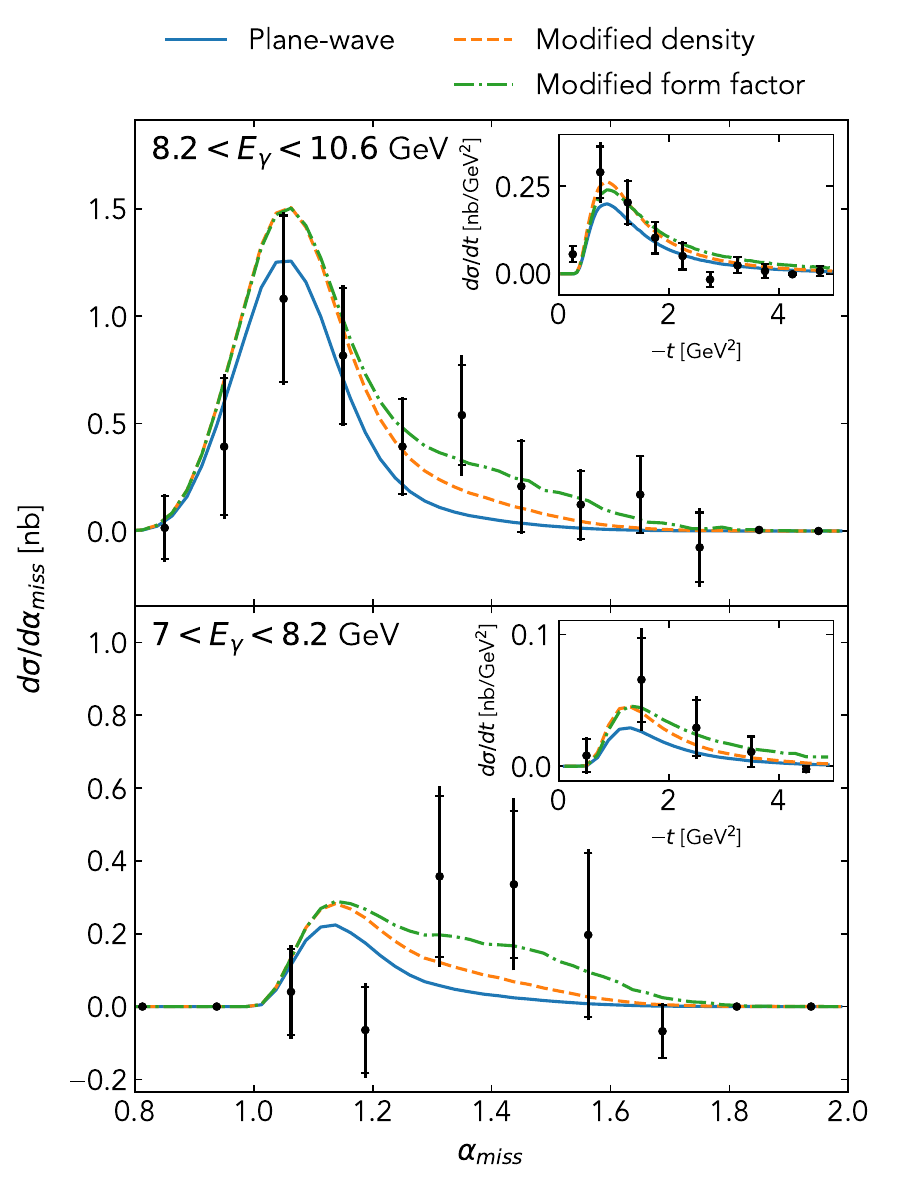}
    \caption{
    Left: Averaged A$(\gamma,J/\psi p)X$ cross section for $^4$He and $^{12}$C as a function of beam photon energy, compared with plane-wave calculations for mean-field (dotted red) and SRC (dashed blue) contributions as well as the total (dot-dashed grey). Data also include a common 23\% normalization uncertainty (not shown).
    Right: Differential A$(\gamma,J/\psi p)X$ cross section as a function of light-cone momentum fraction $\alpha_\text{miss}$, for $^4$He and $^{12}$C, separated into the above-threshold (top) and below-threshold (bottom) energy regions. Insets shows the differential cross section as a function of momentum transfer $|t|$. Measured data (black points) are compared with plane-wave calculations (blue solid line), as well as calculations assuming a modified proton density (orange dashed) and a modified form factor (green dot-dashed). Data also include a common 23\% normalization uncertainty (not shown).
    Figures from Ref.~\cite{pybus2024measurementnearsubthresholdjpsi}.
    }
    \label{fig:subthresholdPRL}
\end{figure}

These data, shown in Fig.~\ref{fig:subthresholdPRL}, provided evidence for enhanced production of subthreshold $J/\psi$ photoproduction as compared with nuclear structure predictions, and suggested that bound protons could have a substantially modified gluon structure.
As a completely novel measurement, this work was selected in Physical Review Letters as an Editors' Suggestion in recognition of its significance in exploring new avenues for understanding the gluonic structure of nuclei.
This established the methodology for the analysis described in this proposal and provides strong motivation for a dedicated, high-luminosity measurement capable of disentangling the impact of the nuclear environment on the gluon structure of bound protons.


\section{Physics Goals}
\label{sec:goals}

Photoproduction of $J/\psi$ from nuclear targets provides the opportunity to perform probes of the gluonic structure of the nucleus, similar to recent studies on the proton~\cite{Ali_2019,adhikari2023measurement,HallCJPsi}. 
Incoherent $J/\psi$ photoproduction from nuclei is sensitive to the fluctuations of gluons within the nucleus~\cite{PhysRevD.18.1696}, as well as the gluonic structure of the bound nucleon.
In photoproduction of $J/\psi$, the photon energy is related to the gluon momentum fraction $x\sim\frac{m_{J/\psi}^2}{2m_NE_\gamma}$.
The study of near-threshold $J/\psi$ photoproduction from nuclei would therefore allow a first search for a gluonic EMC-like effect in the valence region of $x\sim 0.5$.

Of similar interest is the possibility of measuring sub-threshold photoproduction of $J/\psi$~\cite{Hatta_2020}.
In nuclei, the Fermi motion of nucleons enables production of $J/\psi$ at lower photon energies than the production threshold of $E_\gamma\approx8.2$ GeV from the proton.
Such production is predicted to be directly sensitive to the details of nuclear structure.
At sub-threshold energies, the production of $J/\psi$ has a higher contribution from Short-Range Correlations~\cite{Hatta_2020,pybus2024measurementnearsubthresholdjpsi}, enabling a probe of the gluon structure of correlated nucleons.
A detailed scan of nuclear $J/\psi$ photoproduction over the photon energy threshold is at this point only possible at JLab following the 12-GeV upgrade, and this would provide critical insights into the gluon structure of the nucleus.

A high-statistics measurement of $(\gamma,J/\psi\ p)$ photoproduction from $^4$He would enable a scan of the incoherent nuclear photoproduction cross section as a function of photon energy. 
The cross section can be measured with photon energies ranging from 7.5 GeV to the endpoint energy of 12 GeV.
Optimal placement of the coherent photopeak can enable measurement of the cross section even below threshold, where the cross section is expected to be small.

Detecting the proton in such events improves the mass resolution of the $J/\psi$, and additionally enables reconstruction of the initial-state momentum of the proton involved in quasi-free production.
This semi-inclusive measurement will enable detailed study of the reaction mechanisms for sub-threshold production by examining the correlation between the initial nuclear motion and the photon energy of the reaction.
Knowledge of the initial proton momentum can also enable specific study of $J/\psi$ photoproduction from SRC nucleons. 
By comparing $(\gamma,J/\psi p)$ event with low and high initial proton momentum, we can perform the first gluonic probe of SRCs using this process, gaining possible insight into the gluonic content of correlated nucleons.

Photoproduction of $J/\psi$ from nuclei near threshold could be sensitive to a number of exotic effects which may be present in such interactions.
A natural effect to search for would be a gluonic analogue to the EMC effect, which first observed that quark distributions within nuclei are modified relative to those in free nucleons.
Current nuclear PDF fits~\cite{Eskola_2017,2660149} weakly suggest a possible modification of the gluon PDFs within the valence region $0.3\lesssim x\lesssim0.7$, but present data provide minimal constraints to these fits.
Lattice QCD calculations~\cite{PhysRevD.96.094512} also suggest a possible modification of the average gluon momentum fraction within nuclei, but again lack data to substantiate any such predictions.
Photoproduction of $J/\psi$ in the energy range of $7<E_\gamma<12$ GeV corresponds to the valence region, sensitive to an average gluon momentum fraction of $x\sim0.5$, and can provide substantial experimental data to improve our understanding of the nuclear gluon distributions and expose any such effects.

\begin{figure}[t]
    \centering
    \raisebox{0.4\height}{\includegraphics[width = 0.4 \textwidth]{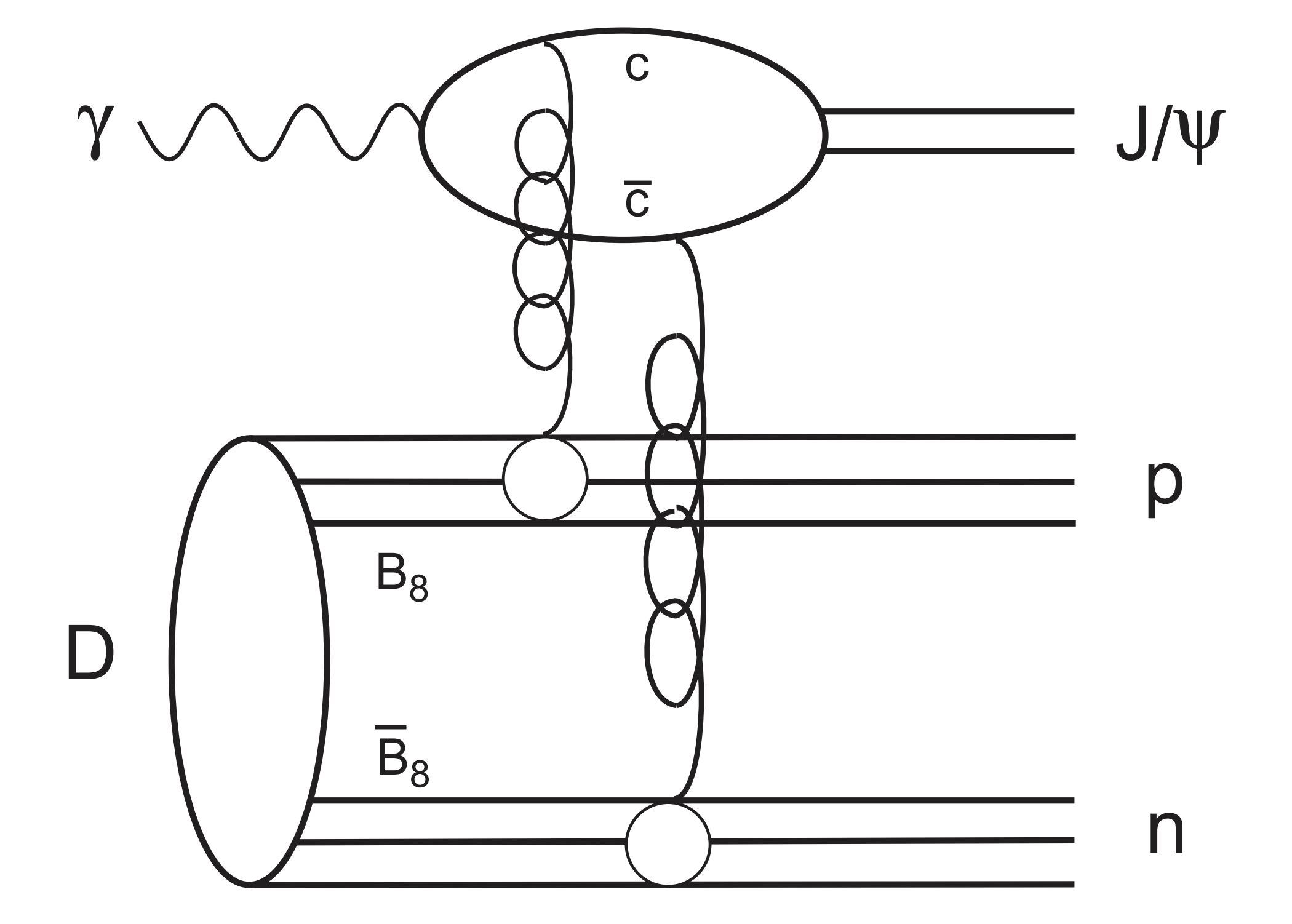}}
    \hspace{8mm}
    \includegraphics[width = 0.4 \textwidth]{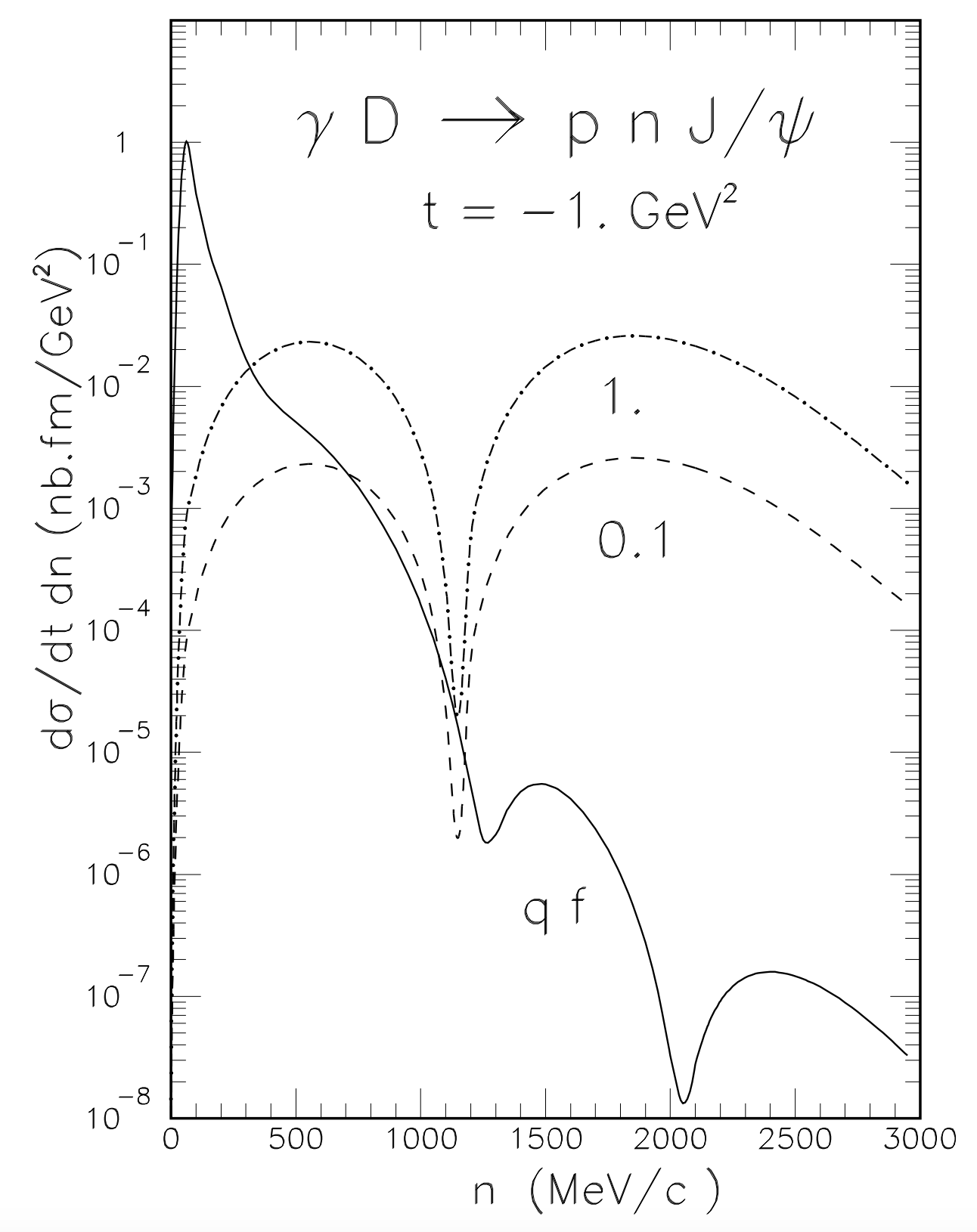}
    \caption{Left: Diagram of possible hidden-color state with coupling to $J/\psi$ photoproduction~\cite{Brodsky_2001}. In this process gluons can be exchanged with both ``color octet'' states within the color-sharing system, resulting in color-neutral nucleons in the final-state.
    Right: Calculated cross section for incoherent $\gamma D\rightarrow J/\psi pn$. The solid line shows the calculation for quasi-free photoproduction from the proton, and the dashed (dotted) lines show the contribution from hidden-color components as 0.1 (1)\% of the nucleus.}
    \label{fig:hidden-color}
\end{figure}

Of similar interest is the possibility of accessing hidden-color states within the nucleus.
These are states in which the nucleus is not well-described by a collection of protons and neutrons as independent color-singlets, but contains collections of 6 or greater quarks which share color between them and cannot be decomposed into constituent hadrons~\cite{6quark,matveev1978quark,PhysRevLett.51.83}.
No definitive experimental evidence for hidden-color states yet exists, but such states are predicted to be present in high-density nuclear states such as SRCs.
Photoproduction of $J/\psi$ has been suggested as a unique probe of such exotic color structures~\cite{LAGET1995397,Brodsky_2001}.
As this process is dominated by the exchange of gluons, it is sensitive to hidden-color initial states by processes such as that shown in Fig.~\ref{fig:hidden-color}(left).
In this diagram, the produced $c\bar c$ pair exchanges gluons with each of the color-octets within a color-sharing pair, producing final-state nucleons and a $J/\psi$ meson from a hidden-color initial configuration.
This process is predicted to result in an enhancement of incoherent $J/\psi$ photoproduction from high-momentum nucleons, as well as an enhancement of the cross section at large $|t|$.
While this process was first theorized for the deuteron, similar coupling would be possible to analogous hidden-color states in $^4$He.
Recent studies suggest that $^4$He in particular may have an substantial hidden-color component~\cite{West:2020rlk}, and the nucleus has both a larger EMC effect and SRC fraction than $^2$H.
$^4$He is therefore an ideal nucleus to search for such a hidden-color signal.

\section{Proposed Measurement}
\label{sec:measurement}

\subsection{Run Conditions}

We list here for the convenience of the reader the proposed run conditions.
The run will be performed with conditions similar to the Short-Range Correlations and Color Transparency (SRC-CT) experiment E12-19-003~\cite{hen2020studying}, which previously measured nuclear targets in Hall D.
The experiment will use the GlueX detector in a standard configuration.

{\bf Target:} 
We propose in this experiment taking data using liquid $^4$He in the otherwise standard GlueX target configuration. 
This target has been previously used in Hall D, in both the SRC-CT experiment~\cite{hen2020studying}, which used similar run conditions to those proposed here, and in the PrimEx Eta experiment~\cite{primex}, which used different run conditions but has successfully run for an extended period at a lower luminosity.
This target will be used for the bulk of the experiment, 80 PAC days.
We will also use a target of liquid $^2$H for calibration purposes and to provide physics comparisons.
This target will be used for only 5 PAC days, and is not the focus of the physics in this proposal.
A liquid $^2$H target was also used previously in the SRC-CT experiment, which measured a similar 4 PAC days of $^2$H data.
As both targets are cryogenic liquid targets, the change of target will provide relatively small overhead when compared to the possible installation of a solid target.

{\bf Trigger:} 
We proposed using a similar trigger configuration to that used in the previous SRC-CT experiment.
This configuration consists of two triggers.
The first trigger is based entirely on the total energy deposition in the calorimeters; in the SRC-CT experiment, this trigger condition was given by $E_{FCAL}+3E_{BCAL}>4.2$ GeV. 
This trigger is optimized for detecting final states with high-energy highly-ionizing particles such as $\gamma$ or $e^\pm$.
The second trigger requires a combination of energy deposition in the calorimeters ($4E_{FCAL} + 5E_{BCAL} > 1$ GeV) as well as a hit in the Start Counter.
This trigger allows detection of final-states with minimally-ionizing particles, such as $\pi^+\pi^-p$ final-states.
These triggers have been demonstrated in the SRC-CT experiment to result in manageable trigger and data rates without overwhelming the data collection capabilities of the Hall.
They have also been seen in data to be efficient for the detection of photoproduction events from 2N-SRCs.
Minor optimizations may be performed to ensure efficient detection of the $(\gamma,J/\psi p)$ final-state.

{\bf Beam and Photon Flux:} 
We request for this experiment the maximum possible electron beam energy of 12 GeV.
This energy is particularly important for the measurement of $J/\psi$ photoproduction, as the cross section for this process grows quickly with the photon energy.
We propose using a standard diamond radiator of $4\times10^{-4}$ radiation lengths in the Tagger Hall, with a placement of the coherent enhancement peak at 8 GeV; the selection of this value is detailed in the following subsection.
The beam current is expected to be 150 nA, the same as used in the SRC-CT experiment. 
We estimate a similar total tagged photon flux, roughly $3.4\times10^7\ \gamma$/s when summing over all tagger energy bins.
This results in an estimated integrated luminosity of $135$ pb$^{-1}\cdot$nucleus ($E_\gamma>6$ GeV) for 80 PAC days on $^4$He, an increase over the SRC-CT data by a factor of 10.
For 5 PAC days of $^2$H, we expect an integrated luminosity of $\sim23$ pb$^{-1}\cdot$nucleus, slightly more than that measured in the SRC-CT data.


\subsubsection{Coherent Photopeak Energy Optimization}

The placement of the coherent peak of the diamond radiator has a significant impact upon the photon flux as a function of $E_\gamma$, and is therefore of greatest relevance when considering the measurement of the $J/\psi$ incoherent cross section.
The placement of the coherent peak was selected in order to maximize our ability to measure the sub-threshold cross section for $J/\psi$ from $^4$He.
A coherent peak at 8 GeV greatly enhances the tagged luminosity below the $J/\psi$ threshold, and results in an estimated $\sim 25$ measured $J/\psi$ events from beam photons with energies $E_\gamma<8$ GeV.

A smaller coherent peak energy of 7.5 GeV was also considered, in order to improve the measurement in the deeper sub-threshold region of $E_\gamma<7.5$ GeV.
It was found that the total number of $J/\psi$ events with $E_\gamma<8$ GeV was reduced to 10 in this case, with only a small relative enhancement to the $E_\gamma<7.5$ GeV bin (which remains below 5 estimated events).
This is primarily an effect of the fact that the Hall D Tagger Hodoscope for $E_\gamma\lesssim7.8$ GeV is only a sampling tagger, and has only a 50\% acceptance for the tagging of beam photons. 
As a result of this, and the very low predicted $J/\psi$ cross section for these photon energies, it is challenging to optimize for a reasonable measurement of the $J/\psi$ deeply-sub-threshold cross section with $E_\gamma<7.5$ GeV.

For completeness we also considered a coherent peak energy at an energy of 9 GeV.
This increases the average beam photon energy and results in an increased total $J/\psi$ yield from $\sim800$ to $\sim1000$.
However, these increases are in photon energy ranges which are already predicted to have relatively high yields; 
the total number of $J/\psi$ events with $E_\gamma<8$ GeV is again reduced to $\sim 10$.

In order to optimize the number of sub-threshold events, we found that a coherent peak energy of 8 GeV resulted in roughly twice as many events with $E_\gamma<8$ GeV than the other two cases considered.
We therefore select this as the optimal coherent peak for mapping out the process $\sigma(\gamma^4\text{He}\rightarrow J/\psi p X)$ as a function of beam photon energy.


It is worth noting that it could be possible to improve the tagging efficiency for lower photon energies by moving the tagger microscope to cover the appropriate tagging region.
This could potentially improve the measurement of sub-threshold $J/\psi$ in the kinematic region $E_\gamma \lesssim7.8$ GeV.
However, the ability to implement this in Hall D without sacrificing tagging efficiency at higher energies requires further study, and this possibility was not factored into either the selection of the coherent peak energy or the estimated event rates.

The estimates in this proposal are performed assuming an electron beam energy of 12 GeV, the maximum energy delivered by CEBAF. 
The experiment could tolerate a small reduction in the energy of the electron beam without major impact on the target measurements.
The primary impact of a reduction of the beam energy to 11.8 GeV would be on the energy-dependent measurement of $J/\psi$; this would result in a 5-10\% reduction in the coherent enhancement and therefore the subthreshold $J/\psi$ event yield, as well as a loss of the highest-energy events. 

\subsection{Final-State Particle Detection}
\label{sec:kin_jpsi}

The quasi-elastic channel $(\gamma,J/\psi p)$ was simulated using a factorized cross section model in the Plane-Wave Impulse Approximation (PWIA):
\begin{equation}
    \frac{d\sigma(\gamma A\rightarrow J/\psi p X)}{dt d^3p_{miss}d E_{miss}} = K\cdot \frac{d\sigma}{dt}(\gamma p\rightarrow J/\psi p)\cdot S(p_{miss},E_{miss})
\end{equation}
where $K$ is a kinematic flux factor, the differential cross section $d\sigma/dt$ for the exclusive process $(\gamma p\rightarrow J/\psi p)$ was taken from a fit to GlueX data~\cite{Ali_2019}, and the spectral function $S(p_{miss},E_{miss{}})$ for Helium was taken from Ref.~\cite{VMC:MFSpecFnc} for the mean-field component and the Generalized Contact Formalism~\cite{Weiss:2018tbu,AxelNature,PYBUS2020135429} for the SRC component.
The generated PWIA events were simulated using the GEANT description of the GlueX detector~\cite{Adhikari_2021}, and were reconstructed using standard GlueX reconstruction software in the same manner as measured data.
In addition, an overall transparency factor of $T = A^{1/3}\approx 0.6$ was assumed.
Total yields were scaled using this factor in order to account for the reabsorption of the final-state proton.

\begin{figure}[t]
    \centering
    \includegraphics[width = 0.32 \textwidth]{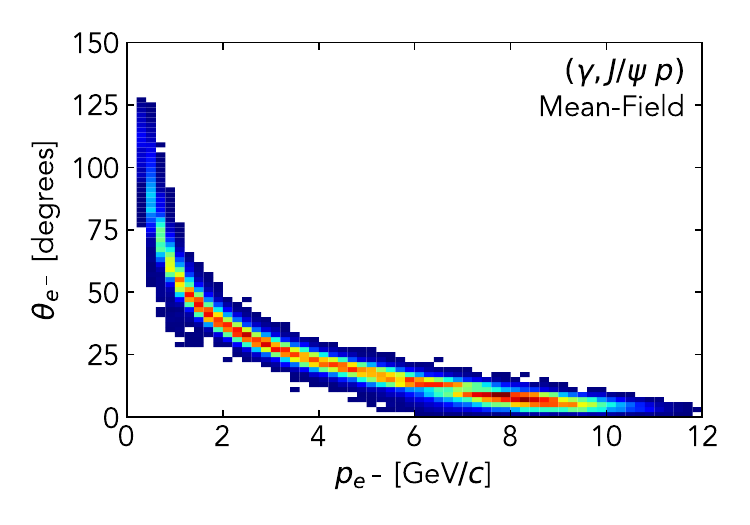}
    \includegraphics[width = 0.32 \textwidth]{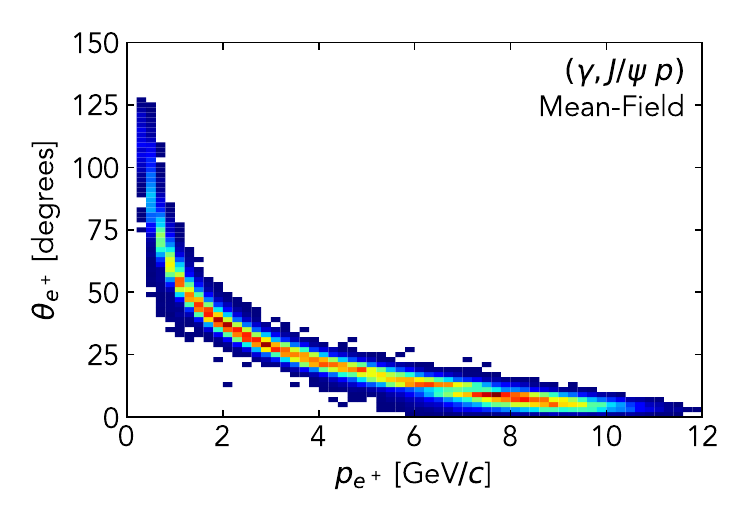}
    \includegraphics[width = 0.32 \textwidth]{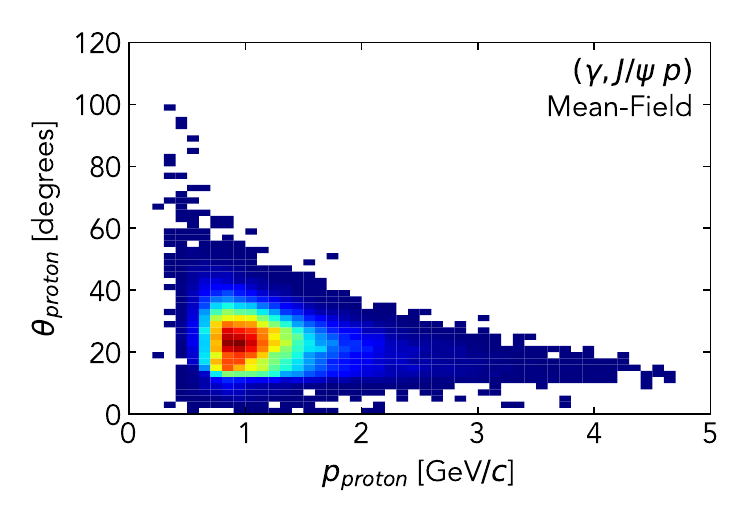}
    \caption{Simulated kinematic distributions for the final-state particles for $(\gamma, J/\psi p)$ production from mean-field protons. The electron (left) and positron (center) have a wide distribution of kinematics but a strong correlation between the momentum and the angle of the lepton. The proton (right) consistently is produced at moderate angles and low momentum.}
    \label{fig:jpsi_MF_kin}
\end{figure}

\begin{figure}[t]
    \centering
    \includegraphics[width = 0.32 \textwidth]{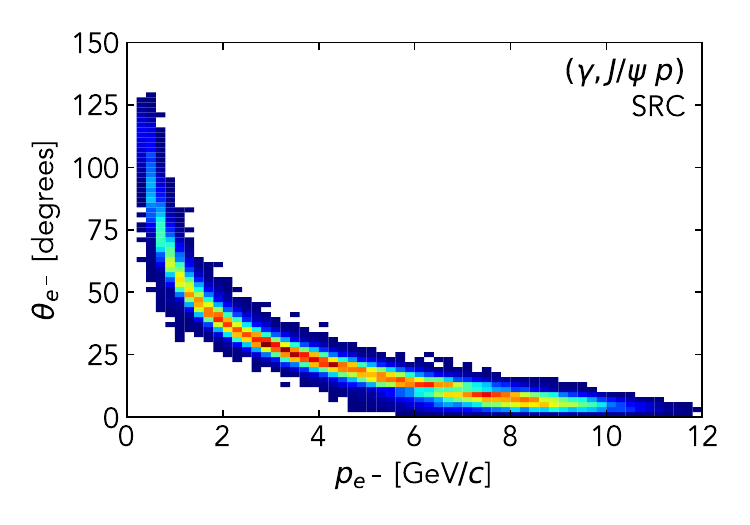}
    \includegraphics[width = 0.32 \textwidth]{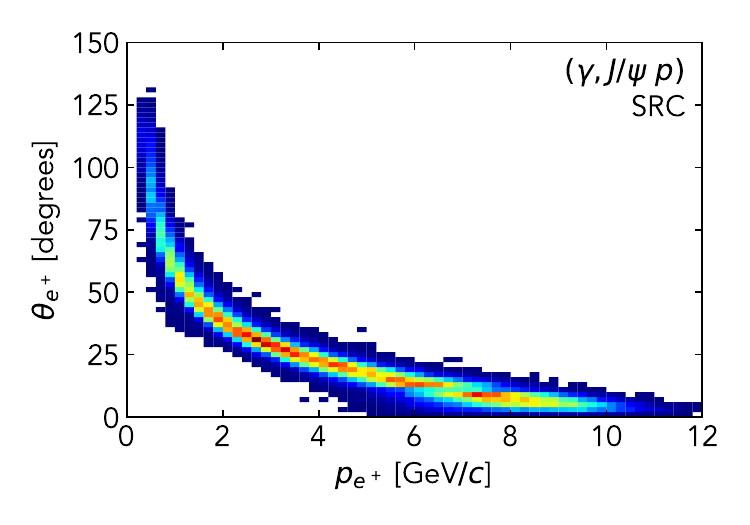}
    \includegraphics[width = 0.32 \textwidth]{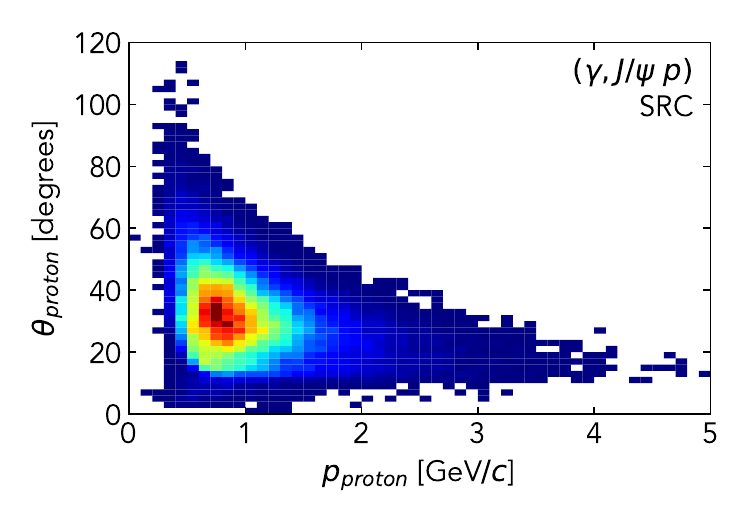}
    \caption{Simulated kinematic distributions for the final-state particles for $(\gamma, J/\psi p)$ production from SRC protons. Kinematics are similar to those events from mean-field protons, shown in Fig.~\ref{fig:jpsi_MF_kin}}
    \label{fig:jpsi_SRC_kin}
\end{figure}

The kinematical distributions of the final-state particles in $(\gamma, J/\psi p)$ events are shown in Fig.~\ref{fig:jpsi_MF_kin} for production from mean-field proton and in Fig.~\ref{fig:jpsi_SRC_kin} for SRC protons. 
For mean-field production, the leptons (electrons and positrons) have a wide kinematic distribution but a strong correlation between the momentum and angle of the particles; these kinematics are strongly controlled by kinematics of the decay $J/\psi\rightarrow e^+e^-$.
The leptons as a result impact in both the Barrel Calorimeter (BCAL) and the Forward Calorimeter (FCAL).
Particle identification for electrons and positrons is primarily possible in the GlueX detector by comparing the energy deposition into the calorimeters with the measured momentum of the charged track; 
for electrons and positrons, these values should have a ratio of $E_{dep}/p_{track}\sim1$.
The protons are consistently produced at low momentum $p_{proton}\sim 1 $ GeV$/c$ and at moderate angles.
The protons therefore primarily impact the BCAL, and additionally frequently have low enough momentum to allow particle identification using $dE/dx$ and time-of-flight.
The kinematics for production from SRC protons are largely similar, with the largest difference being a wider kinematic distribution for the outgoing proton as a result of larger nuclear momentum.

A major consideration in the GlueX detector is successfully resolving the peak of the $J/\psi\rightarrow e^+e^-$ decay.
For this experiment, we will use the light-front method detailed in the analysis of previous data~\cite{pybus2024measurementnearsubthresholdjpsi}.
Full details are given in the Supplemental Materials of the publication, which demonstrates that this choice of mass observable is reconstructible with the resolution of the GlueX spectrometer and remains robust against impact of nuclear motion and final-state effects.

\begin{figure}[t]
    \centering
    \includegraphics[width = 0.4 \textwidth]{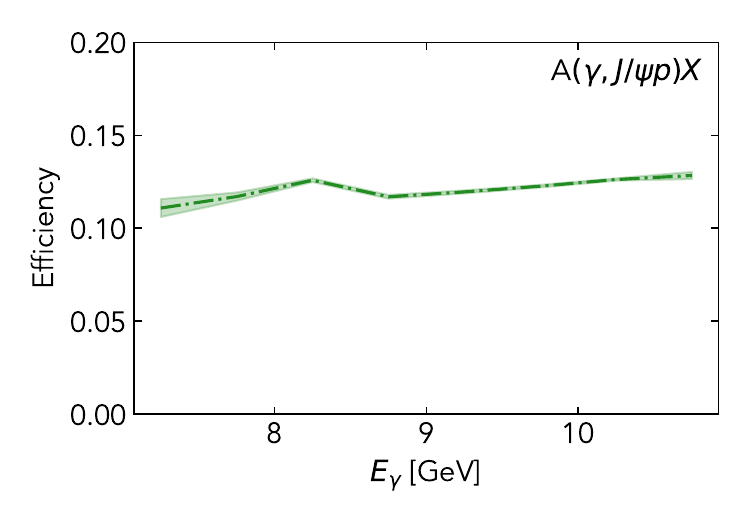}
    \caption{Total efficiency for measuring $A\rightarrow J/\psi p X$ events as a function of $E_\gamma$, calculated using simulation for the current experiment.
    The efficiency is simulated to be between $10-15\%$ and roughly constant with $E_\gamma$.}
    \label{fig:jpsi_eff}
\end{figure}

We also use simulation to estimate the efficiency of detecting $\gamma^4\text{He}\rightarrow J/\psi p X$ events.
Stringent cuts must be placed on data to remove the large backgrounds contributing to an apparently-similar final-state, and it is necessary to quantify the impact of these selection criteria, along with detector efficiency, on the measured signal yield.
Fig.~\ref{fig:jpsi_eff} shows the simulated efficiency for generated $\gamma^4\text{He}\rightarrow J/\psi p X$ events as a function of the beam photon energy. 
This efficiency includes both the inherent detector effects and the impact of particle-identification, fiducial, and energy-balance cuts previously listed. 
We observe that the efficiency is roughly constant as a function of beam energy, and stays between $10-15\%$ over the simulated range.
This efficiency is somewhat smaller than for the exclusive process $\gamma p\rightarrow J/\psi p$ in GlueX, but remains relatively high and sufficient for a differential measurement.

We make an additional note that the GlueX Forward Calorimeter has recently undergone an upgrade to improve its performance~\cite{ASATURYAN2021165683}. 
This upgrade, in addition to improvements in the angular resolution of forward-going showers, is expected to substantially increase the ability of the spectrometer to perform $e/\pi$ separation. 
The improved granularity of the detector will enable greater ability to reject pion backgrounds by examining the shapes of forward-going showers.
Additionally, the installation of a forward Transition Radiation Detector (TRD) is planned~\cite{TRD} in order to further improve $e/\pi$ separation.
This reduced background will improve both point-to-point background uncertainties in the extraction of $J/\psi$ yields as well as normalization uncertainties using Bethe-Heitler background, which is at present contaminated by 2-pion production.

\subsection{Systematic Uncertainties}\label{subsec:SystUncertainties}

We perform estimates of the systematic uncertainties on the measurement of $J/\psi$ photoproduction based on the findings of the analysis of current data~\cite{pybus2024measurementnearsubthresholdjpsi}.
The following sources contribute to the point-to-point uncertainty of the $J/\psi$ cross section extraction:
\begin{itemize}
    \item Bethe-Heitler and 2-pion backgrounds
    \item Cut-dependence
    \item Yield extraction
    \item Bethe-Heitler yield
    \item Efficiency
\end{itemize}

A dominant source of point-to-point uncertainty in the $J/\psi$ measurement is the background in the vicinity of the $J\psi\rightarrow e^+e^-$ decay peak, which is a combination of Bethe-Heitler $e^+e^-$ and misidentified $\pi^+\pi^-$ events.
In the previous experiment this inflated the statistical uncertainties in the measurement by a factor of $\sim1.6-2.0$.
The ongoing upgrades to the GlueX spectrometer will improve forward pion rejection by an expected factor of 10~\cite{TRD,ASATURYAN2021165683}, reducing the background in the vicinity of the $J/\psi$ decay peak by roughly a factor of $4$ and reducing the statistical uncertainty inflation factor from background to $\sim1.25$.

\begin{figure}[t]
    \centering
    \includegraphics[width = 0.4 \textwidth]{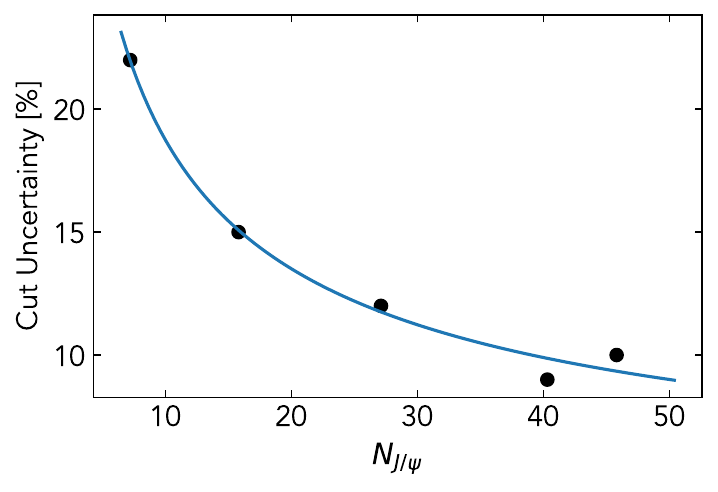}
    \includegraphics[width = 0.4 \textwidth]{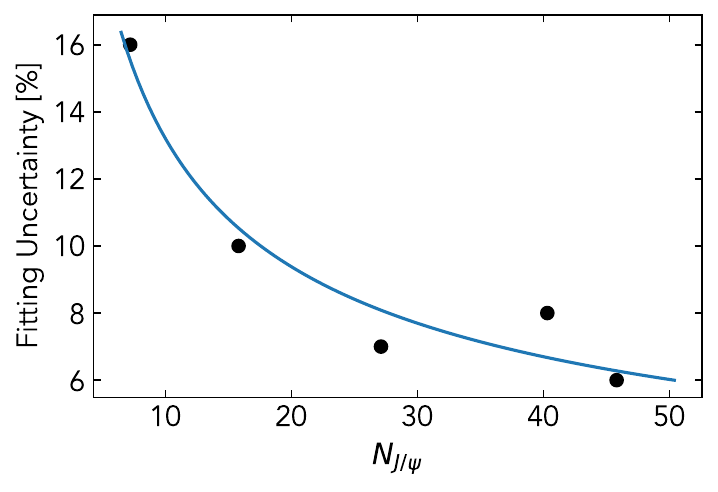}
    \caption{Left: Systematic uncertainty in the extracted $J/\psi$ photoproduction cross section in current $^4$He data resulting from cut-dependence of the measurement, plotted as a function of the measured yield in order to determine the statistical component of the uncertainty. The uncertainties are fit using a $\sqrt{A+B/N_{J/\psi}}$ functional form in order to calculate projected uncertainties in the proposed data. Right: Same as left, but for the systematic uncertainty related to $J/\psi$ yield extraction.}
    \label{fig:unc_vs_N}
\end{figure}

The cut-dependence of the extracted $J/\psi$ cross section is a secondary source of systematic uncertainty, ranging from $10-20\%$ in the previous experiment. 
The primary contribution to this uncertainty is the impact of the restriction on the missing energy of the reaction when assuming photoproduction from a standing proton, a necessary cut in order to remove inelastic backgrounds.
The other contributions to this uncertainty are the cuts used to identify electrons and positrons using energy deposition in the calorimeters.
As this uncertainty is assessed by changing the values of various cuts, it is impacted by statistical fluctuation in the data, following a functional form $\sqrt{A + B/N_{J/\psi}}$; see Figure~\ref{fig:unc_vs_N} (left) for the extraction of this uncertainty from current $^4$He data as a function of the $J/\psi$ yield. 
This uncertainty will be reduced to roughly $10\%$ in most kinematic bins for this reaction; lower-statistics bins will see this uncertainty rise but remain sub-dominant to statistical and background-related uncertainties.

The uncertainty resulting from the extraction of the $J/\psi$ yield ranges from $5-15\%$ in the current data, also subject to statistical uncertainties as well as the level of background. 
The uncertainty has similar behavior to that of the cut-dependence of the measurement, as shown in Figure~\ref{fig:unc_vs_N} (right).
Given the improved statistics of the proposed measurement, this uncertainty is expected to be $5-8\%$ in most kinematic bins. 
In the event that the TRD is implemented in GlueX for this experiment, this source of uncertainty can be anticipated to be reduced further to the percent-level, thanks to a reduction in the background and therefore reduced model-dependence of the $J/\psi$ yield extraction.

Other sources of point-to-point uncertainty are the model-dependence of the extracted $J/\psi$ efficiency as a function of kinematics, estimated be on the order of $1.5\%$ as found in Ref.~\cite{pybus2024measurementnearsubthresholdjpsi}, as well as the point-to-point uncertainty on the Bethe-Heitler yield used to normalize the $J/\psi$ cross section. 
As the previous analysis~\cite{pybus2024measurementnearsubthresholdjpsi} did not perform Bethe-Heitler normalization due to the limited statistics of the data, the previous GlueX analyses~\cite{Ali_2019,adhikari2023measurement} were used to estimate this uncertainty at 5\%.

\begin{figure}[t]
    \centering
    \includegraphics[width = 0.8 \textwidth]{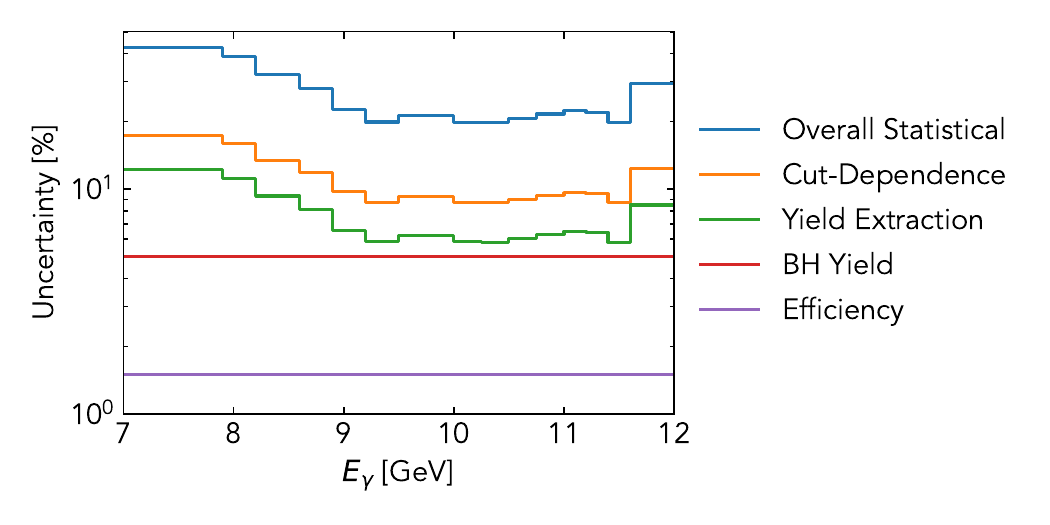}
    \caption{Contributions to the uncertainty on the extracted energy-dependent cross section for $^4$He$(\gamma,J/\psi p)X$ given the proposed binning.
    The total systematic uncertainty not associated with the background is projected to be 10-15\% over most kinematic bins.}
    \label{fig:jpsi_unc}
\end{figure}

Figure~\ref{fig:jpsi_unc} shows the projected sources of point-to-point uncertainty on the energy-dependent cross section extraction for $^4$He$(\gamma,J/\psi p)X$.
The leading source of uncertainty is the statistical uncertainty on the measured yields, followed by the uncertainty on the cut-dependence of the measurement at the roughly 10\% level.
This is followed by the uncertainty on the yield extraction method, the Bethe-Heitler yields, and the efficiency, each of which contribute uncertainty typically below 10\% to the measurement.
Numerical values for these uncertainties are also given in Table~\ref{tab:jpsi_unc}, assuming a typical kinematic bin containing 50 measured $J/\psi$ photoproduction events.
We note that while the systematic uncertainties are anticipated to have a significant impact on this measurement, most of these sources of uncertainty diminish with increasing statistical precision due either to the increase in measured background precision or the decrease in statistical fluctuations when altering elements of the analysis.
As such, the proposed measurement is not projected to be systematically limited in a manner which reduces the effectiveness of increased data quantity; the proposed measurement will be a substantial improvement over the existing data in both statistical and systematic precision.

The dominant source of normalization uncertainty in the previous measurement is the detector efficiency for measuring the $e^+e^-p$ final-state. 
The limited $J/\psi$ statistics prevented normalization of the data using Bethe-Heitler $e^+e^-p$ background, a QED process which is known to good accuracy.
With the improved statistics and 2-pion background rejection of the proposed data given the TRD, the $J/\psi$ measurement will be normalized using Bethe-Heitler as in previous GlueX measurements~\cite{Ali_2019,adhikari2023measurement}.
It is estimated that normalization uncertainty will be reduced below $10\%$ with the improved measurement of Bethe-Heitler after detector upgrades, dominated by radiative effects and non-Bethe-Heitler contributions to the $e^+e^-$ final-state.
Normalization to Bethe-Heitler will also largely cancel uncertainty related to the transparency of the knockout proton, which is the second-largest contribution to the normalization uncertainty in the present data.

\begin{table}[h!]
\centering
\caption{Projected sources of point-to-point uncertainty on extracted $^4$He$(\gamma,J/\psi p)X$ cross sections in a typical kinematic bin expecting a yield of 50 $J/\psi$ photoproduction events.}
\label{tab:jpsi_unc}
\begin{tabular}{|l|l|}
\hline
Source of Uncertainty & Contribution \\ \hline\hline
Statistical           & 14\%                     \\ \hline
Background            & 10\%                      \\ \hline
Cut-dependence        & 9\%                      \\ \hline
Yield extraction      & 2\%                      \\ \hline
Bethe-Heitler         & 5\%                      \\ \hline
Efficiency            & 1.5\%                    \\ \hline\hline
Total                 & 20\%                     \\ \hline
\end{tabular}
\end{table}

\subsection{Expected Rates}

\begin{figure}[t]
    \centering
    \includegraphics[width = 0.4 \textwidth]{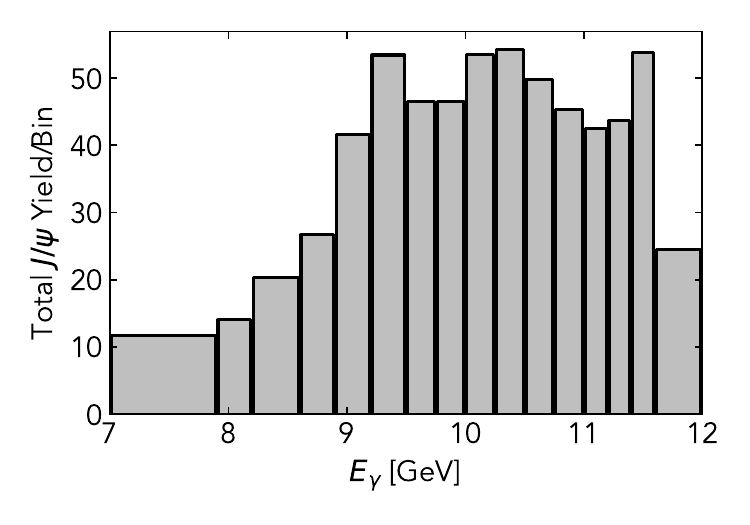}
    \includegraphics[width = 0.4 \textwidth]{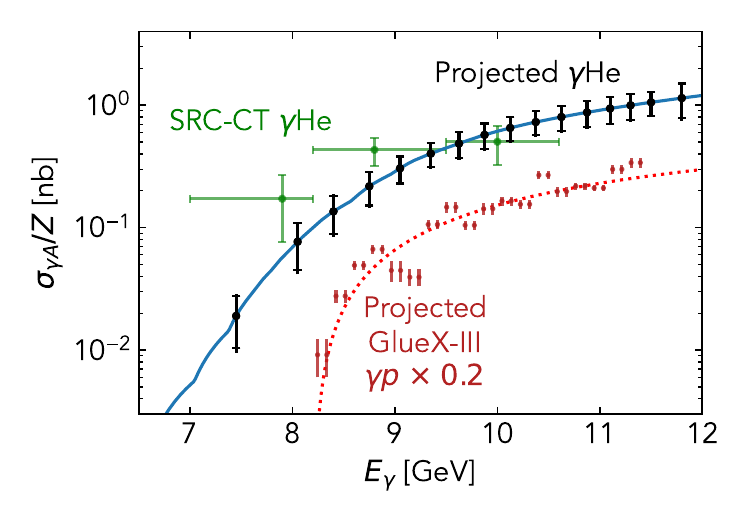}
    \caption{
    (Left): Projected yields for $(\gamma^4\text{He}\rightarrow J/\psi p X)$ as a function of beam photon energy $E_\gamma$.
    Bin sizes were selected to provide a balance between the statistical uncertainties of the points.
    (Right): Projected measurement of $\sigma(\gamma^4\text{He}\rightarrow J/\psi p X)$ as a function of $E_\gamma$. 
    The inner error bars show the estimated statistical uncertainties resulting from the measured $J/\psi\rightarrow e^+e^-$ yield. The outer errors are the estimated total uncertainties, including the contributions from point-to-point systematic uncertainties.
    Not shown is an estimated 10\% overall normalization uncertainty.
    The green points show the results of the SRC-CT experiment's measurement of $^4$He.
    The red curve and points show the $\gamma p\rightarrow J/\psi p$ cross section along with the projected measurement precision of the approved GlueX-III experiment.}
    \label{fig:4HeJPsi}
\end{figure}

The simulations of incoherent $J/\psi$ photoproduction described in Sec.~\ref{sec:kin_jpsi} were used to perform yield estimates for 80 days of $^4$He running.
Fig.~\ref{fig:4HeJPsi} (left) shows the estimated yield of semi-inclusive $(J/\psi p)$ events in bins of beam photon energy $E_\gamma$.
We find that the estimated yields are sufficient to allow a differential measurement in $E_\gamma$, and to provide sufficiently fine binning to map out the cross section over the $J/\psi$ threshold while maintaining adequate statistics in each bin.
Notably, we anticipate a yield of $\sim 25$ subthreshold $J/\psi$ photoproduction events in addition to roughly $800$ higher-energy events.

We also use these yields in bins of $E_\gamma$ to estimate the precision on the total incoherent cross section $\sigma(\gamma^4\text{He}\rightarrow J/\psi p X)$ as a function of $E_\gamma$, as shown in Fig.~\ref{fig:4HeJPsi} (right).
The fractional statistical uncertainties on the cross section are calculated as $1/\sqrt{N}$ for each bin.
The uncertainties resulting from background statistics are estimated to be half the statistical uncertainties.
Other point-to-point systematic uncertainties are estimated to be ~10\%, similar to the previous GlueX study~\cite{Ali_2019} and an improvement over the SRC-CT data, and the overall normalization uncertainty is estimated to be 10\%~\cite{TRD}, an improvement over previous studies thanks to the reduced background in the Bethe-Heitler channel.

\subsection{Proposed Observables}

\begin{figure}[t]
    \centering
    \includegraphics[height = 0.4 \textwidth]{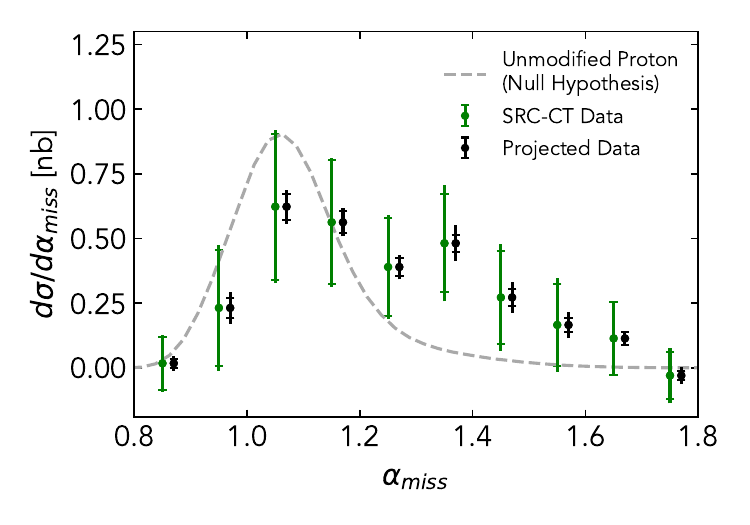}
    \includegraphics[height = 0.4 \textwidth]{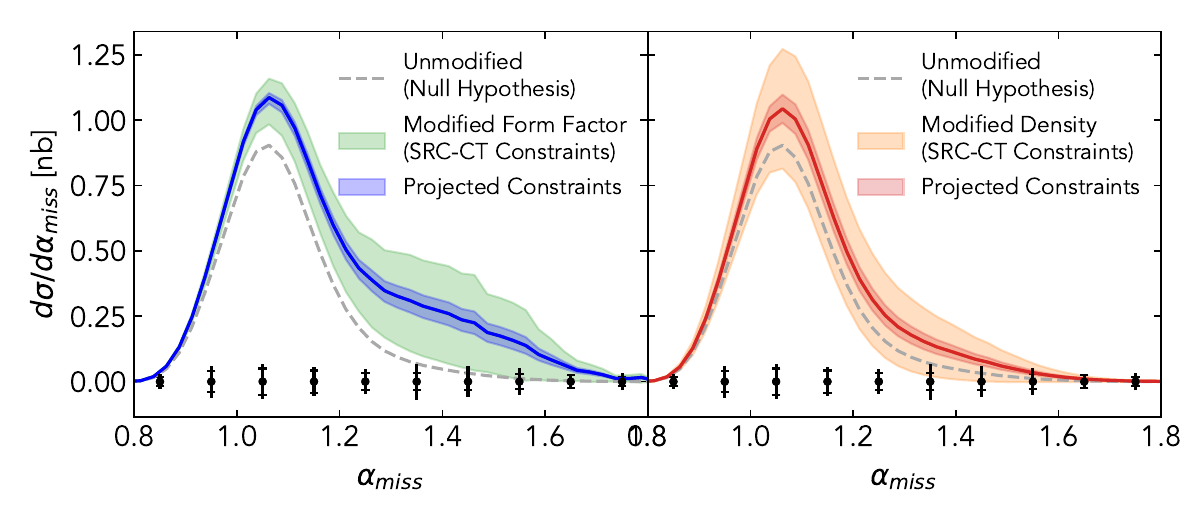}
    \caption{
    (Top): Combined $(\gamma^4\text{He}\rightarrow J/\psi p X)$ and $(\gamma^{12}\text{C}\rightarrow J/\psi p X)$ measured differential cross section as a function of struck nucleon momentum-fraction $\alpha_\text{miss}$ from the SRC-CT analysis (black), compared with the projected differential measurement of the proposed data (red).
    These are compared with plane-wave calculations (grey) assuming an unmodified proton structure in the nucleus, detailed in Ref.~\cite{pybus2024measurementnearsubthresholdjpsi}. 
    (Bottom): Current 90\%-CL constraints on gluon modification in bound protons as a function of $\alpha_{miss}$ are compared with projected constraints for this experiment, for both the form factor (left) and density (right) modification models detailed in Ref.~\cite{pybus2024measurementnearsubthresholdjpsi}.}
    \label{fig:JPsi_alphaM}
\end{figure}

\begin{figure}[t]
    \centering
    \includegraphics[height = 0.4 \textwidth]{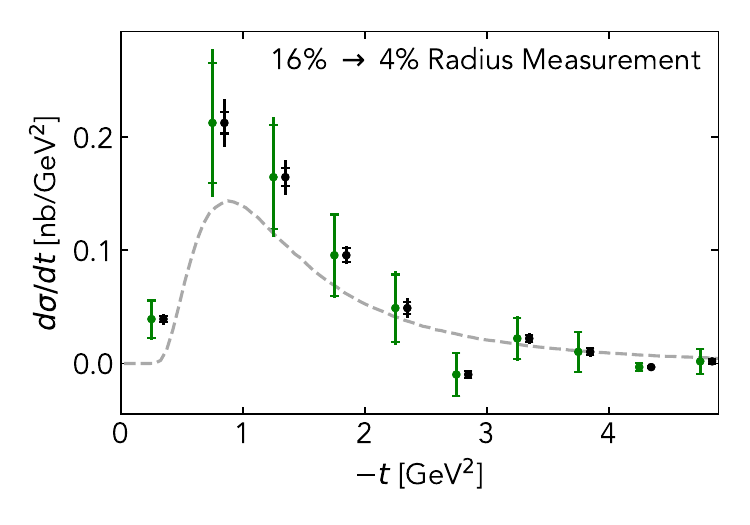}
    \includegraphics[height = 0.4 \textwidth]{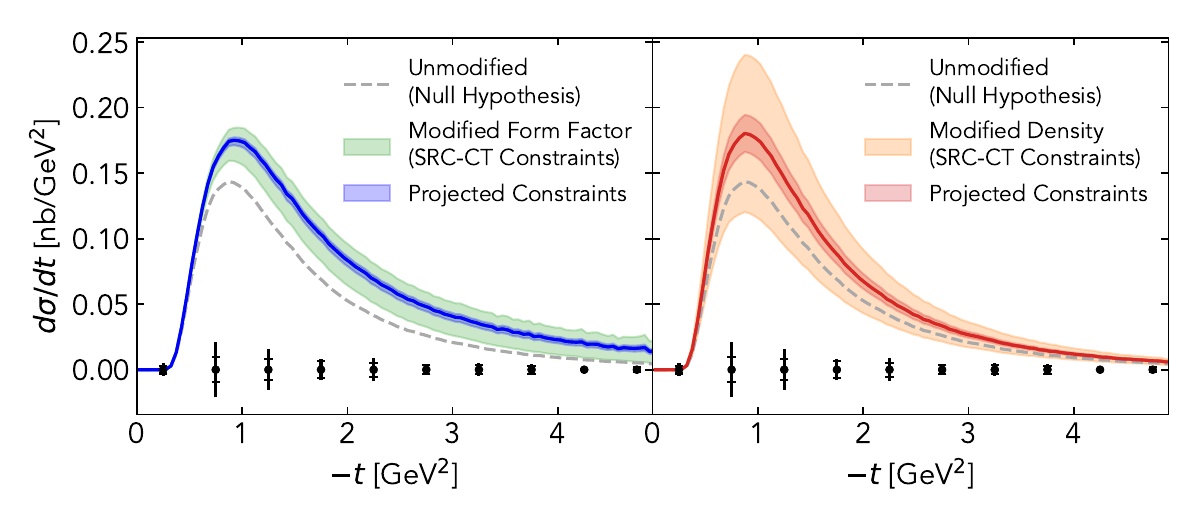}
    \caption{
    Same as Fig.~\ref{fig:JPsi_alphaM} for the differential cross section as a function of $t$. }
    \label{fig:JPsi_t}
\end{figure}

The measurement of $J/\psi$ photoproduction will focus primarily on the semi-inclusive quasi-elastic channel $^4$He$(\gamma,J/\psi p)X$ followed by the decay $J/\psi\rightarrow e^+e^-$. 
This is in large part because the detection of knockout proton substantially improves the resolution of the reconstructed $J/\psi$ mass, allowing greater precision in measuring the event yields.
The target observable is the total cross section for incoherent $J/\psi$ photoproduction as a function of $E_\gamma$, which will have greatest sensitivity to the nuclear gluon distributions as a function of $x$.
The semi-inclusive channel will also enable reconstruction of the momentum-transfer $t$ as well as the initial momentum of the struck proton $p_{miss}$ and its lightcone momentum fraction $\alpha_{miss}$, which will provide more detailed ability to study the underlying production mechanisms and the sensitivity to the initial nuclear state.
Fig.~\ref{fig:JPsi_alphaM} (top) shows the projected measurement of the differential cross section as a function of $\alpha_{miss}$, as compared with existing data and a plane-wave calculation assuming no proton structure modification in the nucleus.
Fig.~\ref{fig:JPsi_alphaM} (bottom) shows an example of the constraints the projected data will provide on different models of gluon modification in bound protons compared to the constraints set by existing data, with the details of the modification models described in Ref.~\cite{pybus2024measurementnearsubthresholdjpsi}.
Fig.~\ref{fig:JPsi_t} shows the same comparisons for the differential cross section as a function of $t$, which is sensitive to the gluonic form factor or radius for bound protons; we note that the proposed data will reduce the uncertainty of the mean in-medium gluonic radius of the proton from 16\% in Ref.~\cite{pybus2024measurementnearsubthresholdjpsi} to 4\%.

Other possible channel of are inclusive $^4$He$(\gamma,J/\psi)X$ events and semi-inclusive $^4$He$(\gamma,J/\psi n)X$, which could provide access to a greater range of reactions, but the ability to measure such channels in GlueX is not yet clear. 
For this reason, semi-inclusive $^4$He$(\gamma,J/\psi p)X$ remains the target channel for this study.

\section{Summary}


We propose a 85-day measurement using the real photon beam in Hall D and the GlueX detector in its standard configuration, including 80 days using a $^4$He target and 5 days using $^2$H, in order to measure nuclear $J/\psi$ photoproduction at and below the energy threshold to much higher precision than present data allow~\cite{pybus2024measurementnearsubthresholdjpsi}.
The high luminosity also for a large number of $J/\psi$ events over a wide energy range, allowing for a detailed probe of high-$x$ gluons in the nucleus not possible at other facilities. 

\bibliographystyle{ieeetr}
\bibliography{references.bib}

\end{document}